\documentclass[twocolumn,aps,prl,floatfix,superscriptaddress]{revtex4-1}
\usepackage{amsmath,amssymb}
\usepackage{graphicx,float}
\usepackage{times,txfonts}
\usepackage{color}
\usepackage{multirow}
\usepackage{bbold}
\usepackage{color}
\usepackage{soul}
\usepackage{hyperref}
\usepackage{times,txfonts}
\usepackage{natbib}
\usepackage{blindtext}

\newcommand{\qo}[1]{``#1''}

\renewcommand{\epsilon}{\varepsilon}
\renewcommand{\phi}{\varphi}

\usepackage{sistyle}
\usepackage{soul}
\definecolor{lightblue}{RGB}{185,210,248}
\sethlcolor{lightblue}

\begin{document}
\title{Polarization shaping for control of nonlinear propagation}
\author{Fr\'ed\'eric Bouchard}
\affiliation{The Max Planck Centre for Extreme and Quantum Photonics, Department of Physics, University of Ottawa, 25 Templeton, Ottawa, Ontario, K1N 6N5 Canada}
\author{Hugo Larocque}
\affiliation{The Max Planck Centre for Extreme and Quantum Photonics, Department of Physics, University of Ottawa, 25 Templeton, Ottawa, Ontario, K1N 6N5 Canada}
\author{Alison M. Yao}
\affiliation{SUPA and Department of Physics, University of Strathclyde, 107 Rottenrow, Glasgow G4 0NG, Scotland, U.K.}
\author{Christopher Travis}
\affiliation{SUPA and Department of Physics, University of Strathclyde, 107 Rottenrow, Glasgow G4 0NG, Scotland, U.K.}
\author{Israel De Leon}
\affiliation{The Max Planck Centre for Extreme and Quantum Photonics, Department of Physics, University of Ottawa, 25 Templeton, Ottawa, Ontario, K1N 6N5 Canada}
\affiliation{School of Engineering and Sciences, Tecnol\'ogico de Monterrey, Monterrey, Nuevo Leon 64849, Mexico.}
\author{Andrea Rubano}
\affiliation{Dipartimento di Fisica, Universit\`a di Napoli Federico II, Compl. Univ. di Monte S. Angelo, via Cintia, 80126 Napoli, Italy}
\author{Ebrahim Karimi}
\email{ekarimi@uottawa.ca}
\affiliation{The Max Planck Centre for Extreme and Quantum Photonics, Department of Physics, University of Ottawa, 25 Templeton, Ottawa, Ontario, K1N 6N5 Canada}
\affiliation{Department of Physics, Institute for Advanced Studies in Basic Sciences, 45137-66731 Zanjan, Iran}
\author{Gian-Luca Oppo}
\affiliation{SUPA and Department of Physics, University of Strathclyde, 107 Rottenrow, Glasgow G4 0NG, Scotland, U.K.}
\author{Robert W. Boyd}
\affiliation{The Max Planck Centre for Extreme and Quantum Photonics, Department of Physics, University of Ottawa, 25 Templeton, Ottawa, Ontario, K1N 6N5 Canada}
\affiliation{Institute of Optics, University of Rochester, Rochester, New York, 14627, USA}
\begin{abstract}
We study the nonlinear optical propagation of two different classes of space-varying polarized light beams -- radially symmetric vector beams and Poincar\'e beams with lemon and star topologies -- in a rubidium vapour cell. Unlike Laguerre-Gauss and other types of beams that experience modulational instabilities, we observe that their propagation is not marked by beam breakup while still exhibiting traits such as nonlinear confinement and self-focusing. Our results suggest that by tailoring the spatial structure of the polarization, the effects of nonlinear propagation can be effectively controlled. These findings provide a novel approach to transport high-power light beams in nonlinear media with controllable distortions to their spatial structure and polarization properties.
\end{abstract}
\pacs{Valid PACS appear here}
\maketitle

\noindent\textit{Introduction:} 
Light beams that can propagate without significant change to their spatial profile are of interest for modern optical technologies and high-power laser systems. Self-trapped light filaments, or spatial solitons, are formed when their spreading due to linear diffraction is carefully balanced by a self-focusing (Kerr) nonlinearity that causes the beam to narrow.  Due to their potential to carry an increased information content, there has been significant interest in the formation of spatial solitons carrying orbital angular momentum (OAM)~\cite{firth:97,desyatnikov:01,musslimani:01,soljacic:01, bigelow:02, bigelow:04}. OAM-carrying beams are characterized by an azimuthal phase dependence of the form $\exp{(i\ell\phi)}$~\cite{allen:92}, where the integer $\ell$ corresponds to the topological charge of the phase singularity present at the beam centre (e.g. see~\cite{yao:11} and references therein). Such beams include the Laguerre-Gauss (LG) modes~\cite{barnett:07}, which are solutions to the paraxial wave equation.

It is well known that in (2+1) dimensions, spatial solitons are unstable in homogeneous Kerr media~\cite{rasmussen:86}. One way to increase their stability is to use a saturable self-focusing medium to prevent the catastrophic collapse due to self-focusing. By using an intensity dependent nonlinear refractive index, $n_2$, it becomes possible to balance out the effects of self-focusing and diffraction. However, even in the case of saturable self-focusing media, it is known that optical beams carrying OAM will fragment into several solitons possessing particle-like attributes~\cite{desyatnikov:01,malomed:05}. In particular, a scalar beam carrying an OAM value of $\ell$ is predicted to break up into $2\ell$ daughter solitons~\cite{firth:97,bigelow:04}.

In comparison with scalar ring solitons that carry a definite non-zero OAM, it has been suggested that stability can be increased by using two beams with opposite OAM to produce a beam with a net zero OAM~\cite{soljacic:01, bigelow:02}. For example, \qo{necklace} (petal) beams, which consist of a scalar superposition of two modes with equal and opposite OAM, have been shown to exhibit quasi-stable propagation in a self-focusing medium although they expand upon propagation~\cite{soljacic:98}. For vectorial superpositions of beams carrying OAM, however, the resulting \textit{vector solitons} have been shown theoretically to exhibit quasi-stable propagation for much larger distances than the corresponding scalar vortex solitons and necklace beams~\cite{desyatnikov:01,bigelow:02}. 

Vector vortex beams are fully correlated solutions to the vector paraxial wave equation that have space-varying polarization distributions. Cylindrical vector (CV) beams are a subclass of vector beams with an axially symmetric polarization profile about the beam's propagation axis~\cite{zhan:09, galvez:12}. Examples, include \textit{radial}, \textit{azimuthal} and  \textit{spiral} polarization distributions. Another class of light beams with non-uniform polarization structure is that of full Poincar\'e beams~\cite{beckley:10, galvez:12}. These are of intrinsic interest because they carry polarization singularities. Such beams typically consist of a superposition of two orthogonally polarized LG modes of different orbital angular momenta~\cite{beckley:10,galvez:12} and thus carry a net value of angular momenta.

In this Letter, we demonstrate, both experimentally and numerically, the stable propagation of space-varying polarized light beams in a saturable self-focusing nonlinear medium. More specifically, our study focuses on vector vortex and Poincar\'e beams traveling through rubidium vapour. We compare the intensity and polarization distributions of the beams at the entrance and exit of the nonlinear cell. This allows us to see how beam-break up is affected both by the net OAM of the beam and by its polarization distribution.\newline

\begin{figure}[t]
	\begin{center}
	\includegraphics[width=0.9\columnwidth]{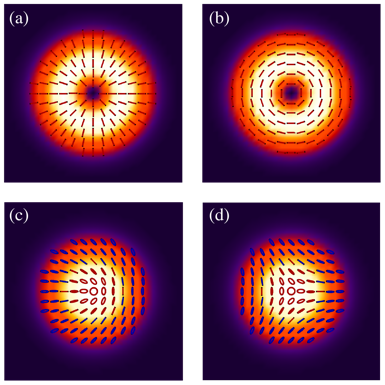}
	\caption[]{(a) Radial and (b) azimuthal vector beams. (c) Lemon and (d) star Poincar\'e beams. Red (blue) ellipses correspond to left (right) circular polarization.}
	\label{fig:beams}
	\end{center}
\end{figure}

\noindent\textit{Theory:} 
Light beams with spatially inhomogeneous polarization distributions can be obtained by a superposition of two spatial transverse modes, $E_1$ and $E_2$, with orthogonal polarizations
\begin{align}\label{eq:superposition}
	\mathbf{E}(r,\phi,z)=E_1(r,\phi,z)\,\mathbf{e}_1+E_2(r,\phi,z)\,\mathbf{e}_2,
\end{align}
where $\mathbf{e}_1$ and $\mathbf{e}_2$ are orthonormal polarization vectors, and $r,\phi,z$ are the cylindrical coordinates. Here we adopt the circular polarization basis, i.e. $\mathbf{e}_1=\mathbf{e}_\text{L}$ and $\mathbf{e}_2=\mathbf{e}_\text{R}$, and the LG basis for the spatial transverse modes~\cite{beckley:10}. 
If the two beams have equal but opposite OAM, the polarization of the beam varies along the azimuthal coordinate and, if the beams are equally weighted, spans the equator of the Poincar\'e sphere. These radially symmetric \emph{vector vortex} beams~\cite{zhan:09} can have polarization distributions that are radial, azimuthal (see Fig. \ref{fig:beams}-(a), (b)) or spiral.
If $E_1$ and $E_2$ carry a zero and a non-zero OAM value, respectively, the resulting full Poincar\'e beam~\cite{beckley:10} has a polarization that varies in both the angular and radial coordinates~\cite{galvez:12} and covers all polarization states on the Poincar\'e sphere~\cite{beckley:10}. The state of elliptic polarization varies with position~\cite{dennis:09} and polarization singularities occur at C-points where the azimuth is not defined and the polarization is circular~\cite{nye:83b}, and along L-lines where the polarization is linear and its handedness is not defined~\cite{nye:83a}. C-points can have three fundamental polarization topologies classified by the index $\eta$ and the number of polarization lines that terminate at the singularity: these topologies are known as ``star'' ($\eta=-1/2$, three lines), ``lemon'' ($\eta=1/2$, one line) and ``monstar'' ($\eta=1/2$, infinitely many, with three straight, lines)~\cite{nye:99, dennis:09}. Examples of lemon and star topologies are shown in Fig.\ref{fig:beams} (c) and (d). 

We simulate propagation through the medium using a two-dimensional nonlinear Schr\"{o}dinger equation with saturable self-focusing nonlinerity, derivable from the two-level model, under the slowly varying envelope and paraxial approximations and normalized to dimensionless quantities, $\rho=r/w_0$ and $\zeta={z}/({2z_R})$, where $w_0$ is the beam waist and $z_R$ is the Rayleigh range of the beam~\cite{firth:97,bigelow:02,desyatnikov:01}. As we are dealing with vector beams, our model consists of two coupled equations that interact through the cross phase modulation (XPM) term characterized by the parameter $\nu$, which takes the value of $2$ for circularly polarized beams~\cite{Agrawal}
\begin{align}
\frac{\partial E_1}{\partial \zeta} = \frac{i}{2}\nabla^2_\bot E_1 +i\mu \; \frac{|E_1|^2+\nu \, |E_2|^2}{1+\sigma(|E_1|^2+ \nu \, |E_2|^2)} E_1, 
	\label{eq:coupled} \\
 \frac{\partial E_2}{\partial \zeta} = \frac{i}{2}\nabla^2_\bot E_2 +i\mu \; \frac{|E_2|^2+\nu \, |E_1|^2}{1+\sigma(|E_2|^2+ \nu \, |E_1|^2)} E_2. \nonumber
\end{align}
The parameters of importance are the nonlinear parameter, $\mu$, and the saturation parameter, $\sigma$, given by:
\begin{equation}
  \mu = \frac{k_0^2 n_2 P_0}{n_0}  \; ; \qquad\quad \sigma = \frac{2P_0}{I_{sat} w_0^2} \, ,
  \label{eq:gammasigma}
\end{equation}
where $k_0$ is the free-space wavenumber, $n_0$ and $n_2$ are the linear and nonlinear refractive indices ($n_2 > 0$ for self-focusing), $I_{sat}$ is the saturation intensity, and $P_0$ is the power of the incident laser beam.
In the simulations reported below, we have selected $\mu=386$, $\sigma=51.7$ that reproduce the experimental configuration of the natural rubidium (Rb) cell. We performed numerical integrations of the propagation Eqs.~(\ref{eq:coupled}),  using the split-step method with fast Fourier transforms and parameters corresponding to the experiments performed.\newline
\begin{figure}[b]
	\begin{center}
	\includegraphics[width=0.8\columnwidth]{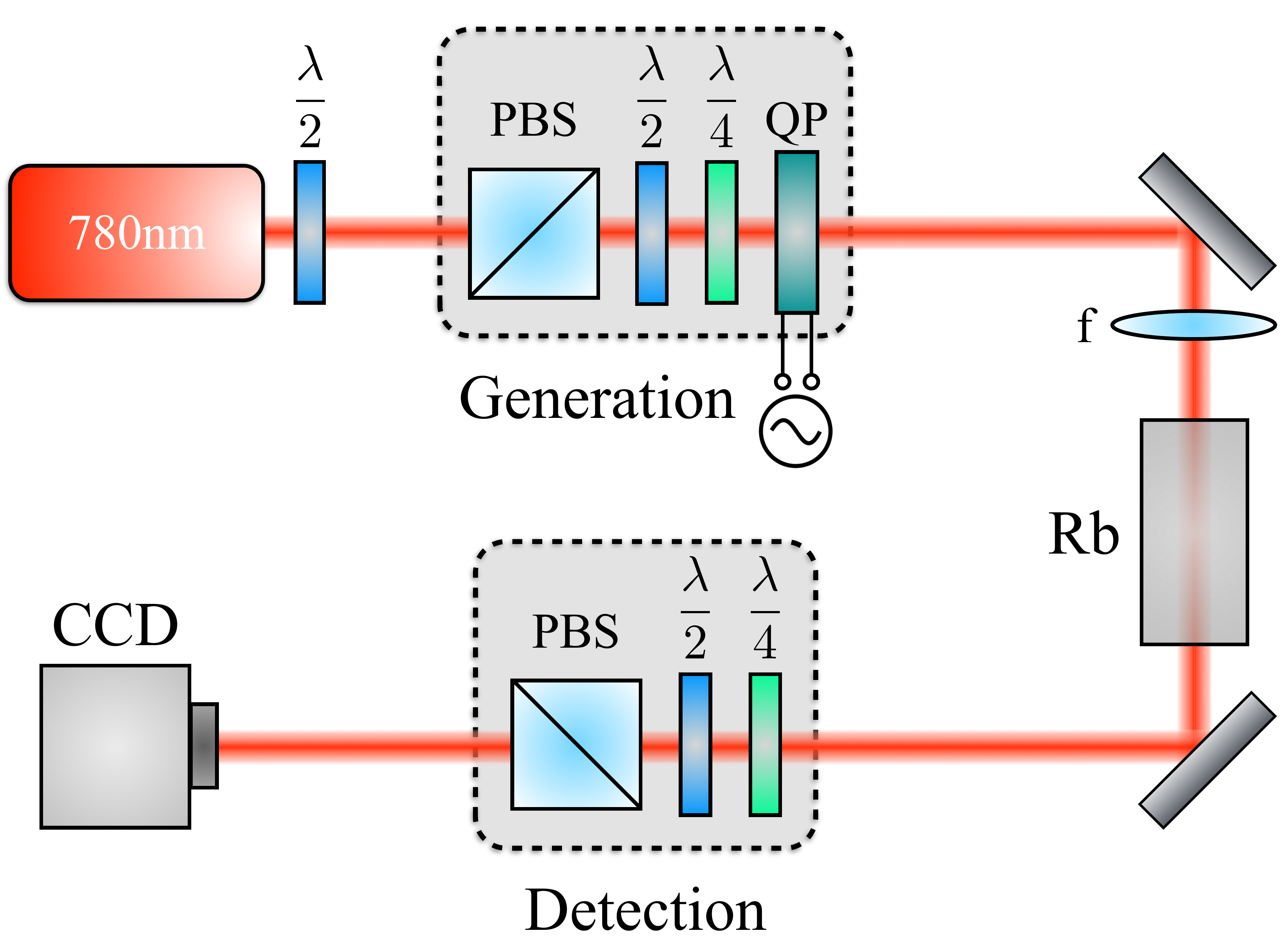}
	\caption[]{Experimental set-up. A CW wavelength-tunable ($760$~nm - $790$~nm) Toptica diode laser is coupled to a polarization-maintaining single-mode optical fiber. The power of the laser beam is adjusted by means of a half-wave ($\lambda/2$) plate followed by a polarizing beam splitter (PBS) in order to reach the saturation threshold ($I_\text{sat}=5$~Wcm${}^{-2}$) for the $D_2$ resonance line of Rb. A combination of a $\lambda/2$ and a $\lambda/4$ plate is used to generate an arbitrary polarization state, which is then sent to a $q$-plate (QP), with a topical charge of $q=\pm 1/2$, to generate vector vortex and Poincar\'e beams. This is then focused ($w_0=60~\mu$m) into a thermally controlled Rb atomic vapour cell by means of a lens ($f=150$~mm). The transmitted beam is analyzed with a combination of a $\lambda/4$ plate, a $\lambda/2$ plate and a PBS (polarization tomography), and attenuated using neutral density filters before being recorded by a CCD camera.}
	\label{fig:exp}
	\end{center}
\end{figure}
\begin{figure*}[t]\centering
	\includegraphics[width=2\columnwidth]{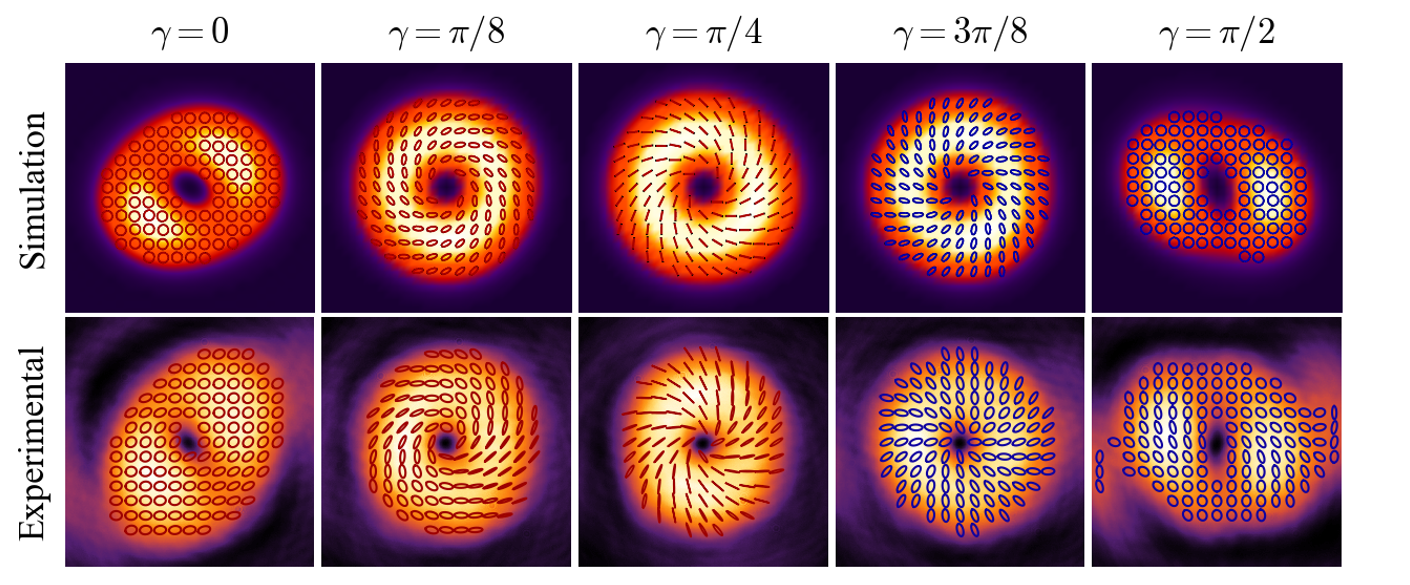}
	\caption[]{Simulated (upper row) and experimentally reconstructed (lower row) intensity and polarization distributions of scalar and vector superposition beams after propagating through the Rb cell. The incident beams are $\cos{(\gamma)}\, \text{LG}_{0,-1}(r,\phi,z)\, \mathbf{e}_\text{L} + \sin{(\gamma)} \,\exp{(i \beta)}\,\text{LG}_{0,1}(r,\phi,z)\,\mathbf{e}_\text{R}$, with $\beta=-\pi/2$ and $\gamma$ specified at the top. Red (blue) ellipses correspond to left (right) circular polarization.} 
	\label{fig:Ratio}
\end{figure*}
\noindent\textit{Experiment:} We use a spatially filtered, linearly polarized, tunable CW single-mode diode laser (Toptica DL pro 780, 760 nm -- 790 nm) together with a set of half- and quarter-wave plates to generate a Gaussian beam with an arbitrary polarization state, $\cos{\left(\theta/2\right)}\,\mathbf{e}_\text{L}+\exp{(i\chi)}\,\sin{\left(\theta/2\right)}\,\mathbf{e}_\text{R}$, where $\theta$ and $\chi$ are set by the orientation of the waveplates. 
This beam is converted into a space-varying polarized light beam using a $q$-plate -- a slab of patterned liquid crystal -- that couples optical spin to OAM~\cite{marrucci:06}. The \emph{unitary} action of a $q$-plate in the circular polarization basis is described by
\begin{eqnarray}
\label{eq:qPlateUnitary}
\hat{\text{U}}_q\cdot\left[\begin{matrix} 
	\mathbf{e}_\text{L} \\
	\mathbf{e}_\text{R} \\
	\end{matrix} \right] = \cos{\left(\frac{\delta}{2} \right)}\left[\begin{matrix}
	\mathbf{e}_\text{L} \\
	\mathbf{e}_\text{R} \\
	\end{matrix} \right]+ i \sin{\left(\frac{\delta}{2}\right)} \left[\begin{matrix}
	\mathbf{e}_\text{R} e^{+2 i \left( q \phi + \alpha_0 \right)} \\
	\mathbf{e}_\text{L} e^{-2 i \left( q \phi + \alpha_0 \right)} \\
	\end{matrix} \right],
\end{eqnarray}
where $q$ is a half-integer number corresponding to the topological charge of the liquid crystal pattern, $\alpha_0$ is the azimuthal orientation of the liquid crystal elements at $\phi=0$ (laboratory frame) and $\delta$ is the optical retardation of the $q$-plate. The parameter $\delta$ can be experimentally adjusted by applying an electric field onto the plate in such a way that the resulting optical retardation corresponds to a half ($\delta=\pi$) or quarter ($\delta=\pi/2$) wavelength~\cite{slussarenko:11}. 
The generated beam is then focused by a $150$-mm-focal-length lens into a $9$-cm-long cell containing Rb atomic vapour. A detailed depiction of this experimental apparatus is provided in Fig.~\ref{fig:exp}. 
The intensity of the incident beam and the temperature of the atomic vapour are set so that the medium exhibits saturable Kerr nonlinearities (powers in the vicinity of $7.44$~mW and a temperature of $95^\circ$C). In order to observe beam breakup of OAM-carrying beams, the laser's output wavelength was tuned near the $D_2$ transition line of Rb. Moreover, the laser was blue-detuned by less than $0.7$~GHz from the $5S_{1/2}(F=3) \rightarrow 5 P_{3/2} (F=4)$ hyperfine transition.

The beam exiting the Rb cell is then imaged using a lens with a focal length of $200$~mm (not shown in the experimental setup) and its polarization distribution is reconstructed using polarization tomography. The tomography is performed using an appropriate sequence of a quarter-wave plate, a half-wave plate, a polarizer and a spatially resolving detector (CCD camera) set at an exposure time of $0.1$~s. 

In order to generate vector vortex beams (radial, azimuthal and spiral), a linearly polarized Gaussian beam is sent to a tuned ($\delta=\pi$) $q$-plate of topological charge $q=1/2$~\cite{cardano:12}. Apart from a global phase, the generated beam will be given by $\left(\text{LG}_{0,-1}(r,\phi,z)\, \mathbf{e}_\text{L} + \exp{\left(i \beta \right)}\,\text{LG}_{0,1}(r,\phi,z)\,\mathbf{e}_\text{R} \right)/\sqrt{2}$, where $\beta \, \equiv 4 \alpha_0$ depends on the orientation of the $q$-plate with respect to the input polarization. The generated beams correspond to radial ($\beta=0$), azimuthal ($\beta=\pi$) and spiral ($\beta=\pm \pi/2$) vector vortex beams. 

To generate the lemon and star Poincar\'e beams, a circularly polarized Gaussian beam is sent to a perfectly detuned ($\delta=\pi/2$)  $q$-plate with $q=1/2$ and $q=-1/2$, respectively~\cite{cardano:13}. Here, $\beta=2\alpha_0$ and does not affect the polarization topology but does cause a rotation in the polarization pattern. Thus, we choose $\beta = 0$, resulting in an output beam of the form $(\text{LG}_{0,0}(r,\phi,z)\, \mathbf{e}_\text{L} +\text{LG}_{0,2q}(r,\phi,z)\, \mathbf{e}_\text{R})/\sqrt{2}$, again omitting any global phase factors. Note also that monstar topologies cannot be readily generated in the laboratory. This scheme allows us to switch between different structured light beams without altering their intensities. \newline 

\noindent\textit{Analysis: }
It is well-known that azimuthal modulational instabilities associated with the helical phase structure found in beams carrying OAM result in their filamentation as they propagate through self-focusing media~\cite{firth:97,desyatnikov:01,bigelow:02}. It has been shown both analytically and numerically, however, that the dominant low-frequency perturbations that typically disrupt ring solitons are inhibited for vector solitons with no net OAM~\cite{desyatnikov:01,bigelow:02}. 
Here we confirm this experimentally and numerically by propagating a beam of the form $\cos{\gamma}\;\text{LG}_{0,-1}(r,\phi,z)\; \mathbf{e}_\text{L}+\sin{\gamma}\; \exp{(i\beta)}\;\text{LG}_{0,1}(r,\phi,z)\;\mathbf{e}_\text{R}$ through a self-focusing medium. 
When $\gamma = 0,\pi/2$, the beam is a scalar LG mode of $\ell=-1 , +1$ with left, right-hand circular polarization. For $\gamma = \pi/4$ we have 
a vector vortex beam with linear polarization and for $\gamma = \pi/8, 3 \pi/8$, a vector vortex beam of left, right-hand elliptical polarization, respectively, following the topology of the vector beam with $\gamma = \pi/4$. Note that each beam has the same total intensity. 

A comparison of the experimental results with the simulations based on Eqs.~(\ref{eq:coupled}) is shown in Fig.~\ref{fig:Ratio} for a vector beam defined by $\beta = -\pi/2$. From our results we can see that the nonlinearity counterbalances diffraction up until the point at which the beams fragment. 
As expected, we see that the scalar OAM-carrying beams ($\gamma=0, \, \pi/2$) are starting to break-up at the exit of the nonlinear cell. The vector beam ($\gamma = \pi/4$), on the other hand, seems almost unperturbed, both in terms of its amplitude and polarization distribution.
Indeed, we can see numerically that the fragmentation point occurs much later for vector beams, $\sim 16$~cm, than for scalar beams, $\sim 9$~cm. 
For the  elliptically polarized vector beams ($\gamma = \pi/8, 3 \pi/8$) the stability length is between the cases of scalar and linear vector vortex beams. In this case the effect of the nonlinear propagation is also evident in the change of polarization distributions: the initial spiral distribution has now become almost azimuthal or radial, respectively (see Fig.~\ref{fig:beams}). This suggests that vector beams allow for greater control over the nonlinearity. 

In spite of the approximations used in the derivation of Eqs.~(\ref{eq:coupled}), small differences in polarization distributions between the experiment and the numerical simulation are only really evident in the biased mode cases of $\gamma = \pi/8, 3 \pi/8$, thus demonstrating the robustness of the vector vortex beam ($\gamma = \pi/4$).

It has been proposed that the increase in stability seen in vector beams is due to the fact that they carry no net OAM~\cite{bigelow:02}. We therefore repeated our analysis using lemon and star polarization topologies which have a net OAM. 
\begin{figure}[t]\centering
	\includegraphics[width=\columnwidth]{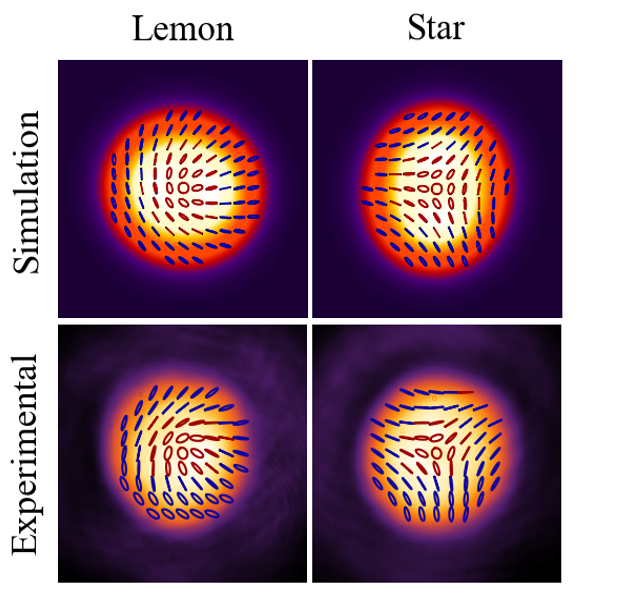}
	\caption[]{\label{fig:Poincare} Intensity and polarization distributions of lemon and star topologies after propagating through the Rb cell, numerical (upper row) and experimental (lower row). Red (blue) ellipses correspond to left (right) circular polarization.
}
\end{figure}
The experimental and numerical results are shown in Fig.~\ref{fig:Poincare} where we have plotted the intensity and the polarization distributions for the lemon and star topologies after propagating through the Rb cell. These results again show that beam break-up has been inhibited when compared to scalar OAM-carrying beams. This result demonstrates that the increased stability is not simply due to a net OAM of zero. As in the case of linear vector vortex beams ($\gamma=\pi/4$) above, the polarization distribution remains unaltered after nonlinear propagation and even preserves the number and kind of polarization singularities. \newline

\noindent\textit{Conclusion:}
We have compared the propagation of scalar OAM-carrying beams with two different classes of beams with non-uniform transverse polarization distributions in a saturable self-focusing nonlinear medium of Rb vapour. With respect to scalar vortex beams (LG modes), we found that beam break up can be inhibited while nonlinear confinement, self-focusing and polarization distributions are not altered for specific cases of non-uniform spatial polarization, both with and without net OAM. This suggests that the spatial structure of the polarization plays an important role in preventing beam fragmentation. These findings provide a novel approach to transport high-power light beams in nonlinear media with controllable distortions to their spatial structure and polarization properties.

\noindent\textit{Acknowledgment:} F.B., H.L., E.K. and R.W.B. acknowledge the support of the Canada Excellence Research Chairs (CERC) program. A. R. acknowledges funding from the European Union (FP7-PEOPLE-2012-CIG, PCIG12-GA-2012-326499-FOXIDUET). E.K. acknowledges the support of the Canada Research Chairs (CRC) program and Canada Foundation for Innovation (CFI).

\end{document}